\documentclass[12pt,epsf]{article}
\usepackage{graphicx, amsmath}

\begin{document}
\sloppy
\begin{flushright}{SIT-HEP/TM-48}
\end{flushright}
\vskip 1.5 truecm
\centerline{\large{\bf Non-standard kinetic term as}}
\centerline{\large{\bf a natural source of non-Gaussianity}}
\vskip .75 truecm

\centerline{\bf Tomohiro Matsuda\footnote{matsuda@sit.ac.jp}}
\vskip .4 truecm
\centerline {\it Laboratory of Physics, Saitama Institute of Technology,}
\centerline {\it Fusaiji, Okabe-machi, Saitama 369-0293, 
Japan}
\vskip 1. truecm

\makeatletter
\@addtoreset{equation}{section}
\def\theequation{\thesection.\arabic{equation}}
\makeatother
\vskip 1. truecm

\begin{abstract}
\hspace*{\parindent}
We consider reheating after inflation with a non-standard kinetic
 term. 
We show that the difference in the kinetic term in different
 Hubble patches inherited from the long-wavelength moduli
 inhomogeneities may generate a significant level of non-Gaussianity
 after inflation.
\end{abstract}

\newpage
\section{Introduction}
Inflation is the main paradigm for understanding the initial conditions
for the cosmological perturbations in the early Universe. 
The observation of the temperature anisotropy of the cosmic microwave
background (CMB) supports the standard inflation scenario that predicts
a scale-invariant and Gaussian spectrum, except for some anomalies. 
An obvious signature of such an anomaly is the observation of a spectrum
index $n\ne 1$, which suggests that the spectrum is not exactly
scale-invariant\cite{EU-book}. 
The observation of $n\ne 1$ is important because it
distinguishes several inflationary models. 
In this paper, we consider another signature of inflation:
non-Gaussianity in the spectrum\cite{Bartolo-text, NG-obs, rod-loop}.
In fact, some inflationary models predict some level of non-Gaussianity
in the spectrum. 
For example, generation of non-Gaussian perturbations may occur (1)
during inflation by a step-like 
potential\cite{Step-inflation}, (2) during inflation by a kinetic
term\cite{Nonlocal_DBI_inflation}, (3) during inflation by a modulated
velocity\cite{Modulated1, Modulated-Kin}, (4) at the end of
inflation\cite{End-Modulated, End-multi, End-multi-mat}, 
(5) after inflation at (p)reheating\cite{IH-PR, IH-R, IH-string, 
Preheat-ng}, and (6) long after inflation\cite{curvaton-dynamics, unco-NG,
curvaton-mazumdar, curvaton-matsuda}.
Except for cases (1) and (2), long-wavelength inhomogeneities of a light
field give rise to non-Gaussianity. 
It is important to consider models of inflation in which some level of
non-Gaussianity follows when a curvature perturbation is generated. 
It would also be interesting to consider possibilities for adding
non-Gaussian perturbations at some event in the early Universe, if it
appears as a natural consequence of basic properties of the model. 
In this paper, we consider reheating after inflation with a non-standard
kinetic term. 
Reheating is a common feature of the inflationary
scenario, and a non-standard kinetic term arises naturally in some
string inspired models and also in other extended models of general
gravity.

\section{The model}
We consider preheating in a theory of two interacting scalar
fields $\{\phi, \chi \}$ with moduli $\sigma$ with non-canonical
kinetic term:
\begin{eqnarray}
\label{hybrid-pot}
S&=&\int d^4 x \sqrt{-g} \left[\frac{1}{2}M_p^2 {\cal R}
-\frac{\omega(\sigma)}{2}\left(\partial_\mu\phi\right)
\left(\partial^{\mu}\phi\right)\right.\nonumber \\ 
&&\left.-\frac{1}{2}\left(\partial_\mu\chi\right)
\left(\partial^{\mu}\chi\right)
-\frac{1}{2}\left(\partial_\mu\sigma\right)
\left(\partial^{\mu}\sigma\right)
-V(\phi,\chi,\sigma),
\right]
\end{eqnarray}
where $M_p$ is the reduced Planck mass and the potential
$V(\phi,\chi,\sigma)$ is given by
\begin{equation}
V(\phi,\chi,\sigma)=\frac{1}{2}m_\phi \phi^2+\frac{1}{2}m_\chi \chi^2
+\frac{g^2}{2}\phi^2\chi^2 +W(\sigma).
\end{equation}
We assume for simplicity that $\phi$ is the inflaton field, which starts
oscillating about the minimum after inflation, and $\chi$ is the preheat
field whose perturbations $\delta \chi$ grow rapidly near the enhanced
symmetric point (ESP). 
Inflation ends and oscillation starts when $\phi=\phi_0$, which gives
the initial amplitude of the $\phi$-oscillation. 
The background fields evolve according to the equations 
\begin{eqnarray}
\ddot{\phi}+3H \dot{\phi}+\frac{(m_\phi^2 +g^2\chi^2)\phi}{\omega}
+\frac{\omega_\sigma}{\omega}\dot{\phi}\dot{\sigma} &=&0\\
\ddot{\chi}+3H \dot{\phi}+(m_\chi^2+g^2\phi^2) \chi &=&0\\
\ddot{\sigma}+3H \dot{\sigma}+W_\sigma-\frac{1}{2}\omega_\sigma 
\dot{\phi}^2&=&0,
\end{eqnarray}
where the subscripts for $\omega$ and $W$ indicate derivatives with
respect to the fields. 
The Hubble parameter is given by
\begin{equation}
H^2\equiv \frac{1}{3M_p^2} \left[\frac{1}{2}\omega \dot{\phi}^2
+\frac{1}{2}\dot{\chi}^2+\frac{1}{2}\dot{\sigma}^2 +V
\right].
\end{equation}
For the instant preheating scenario, the velocity of the oscillating
field at the bottom of the potential determines the number density of
the preheat field. 
For a standard kinetic term (i.e., for $\omega\equiv 1$), the velocity
of the oscillating field is given by
\begin{equation}
\dot{\phi}_{max}\simeq m_\phi \phi_0,
\end{equation}
while for a non-standard kinetic term, it is given by
\begin{equation}
\label{dmax}
\dot{\phi}_{max}\simeq \frac{m_\phi \phi_0}{\sqrt{\omega}}.
\end{equation}
Therefore, $\dot{\phi}_{max}$ at the bottom of the potential is
different in different Hubble patches if the perturbation $\delta \omega$ 
is inherited from the long-wavelength moduli inhomogeneities 
$\delta \sigma$.
The condition for the perturbations of $\sigma$ to cross the horizon 
during inflation is
\begin{equation}
\label{eta-eff}
\eta^{eff}_{\sigma}=M_p^2\left[
\frac{W_{\sigma\sigma}-\omega_{\sigma\sigma}\dot{\phi}^2/2}{V}
\right] 
\simeq 
M_p^2\left[
\frac{W_{\sigma\sigma}}{V}
-\frac{1}{2V}\frac{\omega_{\sigma\sigma}}{\omega^2}
\left(\frac{m^2_{\phi}\phi}{3H} \right)^2
\right] \sim -\eta_{\phi}^2\frac{\omega_{\sigma\sigma}\phi^2}{\omega^2}
\ll 1,
\end{equation}
where $ W_{\sigma\sigma}/3H^2\ll 1$ is assumed.
$\sigma$ moves slowly during inflation if the following slow-roll
condition is satisfied: 
\begin{equation}
\epsilon^{eff}_{\sigma}=\frac{M_p^2}{2}\left[
\frac{W_\sigma-\omega_\sigma\dot{\phi}^2/2}{V}
\right]^2 \sim \eta_\phi^4 \omega_\sigma^2 \phi^2 \left(
\frac{\phi^2}{M_p^2}\right)\ll 1,
\end{equation}
where $\frac{M_p^2}{2} \left(\frac{W_\sigma}{V}\right)^2\ll 1$ is assumed.
To understand these conditions, we consider a specific choice of
$\omega$:
\begin{equation}
\label{choice-omega}
\omega = e^{\alpha \frac{\sigma^2}{M_*^2}},
\end{equation}
where $\alpha$ is a dimensionless constant and $M_*$ is the cut-off scale
of the effective theory.
We find from Eq.(\ref{eta-eff}) that
\begin{equation}
-4\alpha^2 \frac{\sigma^2}{M_*^2}-2\alpha \ll 
\frac{M_*^2 \omega}{\eta_\phi^2 \phi^2}.
\end{equation}
Note that the field $\sigma$ has negative $\eta_\sigma^{eff}$ for
$\alpha>0$.
The condition from the effective $\epsilon$-parameter leads to a similar
condition 
\begin{equation}
4\alpha^2 \frac{\sigma^2}{M_*^2}
\ll \frac{M_*^2M_p^2}{\omega^2\eta_{\phi}^4 \phi^4}.
\end{equation}
Following these conditions, we assume a natural condition 
$\sigma \ll M_*$ and consider a slow-rolling $\sigma$ field during
inflation. 

Considering the reheating process after inflation, the inflaton field
must finally decay into the Standard-Model (SM) particles. 
Here we consider the primary decay process of the inflaton field:
$\phi \rightarrow \chi$. 
Since the interaction depends on the values of the fields $\phi$ and
$\chi$, the background 
field trajectories will be very sensitive to the initial conditions and
the non-perturbative effects of the preheating process, 
which means that the general evaluation of the cosmological parameters
typically requires numerical calculations. 
In this paper, we consider the instant preheating
scenario\cite{instant-PR} so that an analytic estimate can be 
made for the non-linear parameter for the CMB spectrum. 
Applying the results of Ref.\cite{instant-PR}, the comoving number
density of the preheat $\chi$
particles produced at the ESP during the first half-oscillation is 
\begin{equation}
\label{preheat-n}
n_\chi=\frac{(g|\dot{\phi}_{max}|)^{3/2}}{8\pi}
\exp\left\{-\frac{\pi g |\phi_{min}|^2}{|\dot{\phi}_{max}|}
\right\}\simeq \frac{(g|\dot{\phi}_{max}|)^{3/2}}{8\pi}.
\end{equation}
Due to the interaction term, the effective mass of the preheat field
increases as the oscillating field moves away from the bottom of the
potential. 
Thus, the preheat particles produced soon acquire large mass
and decay into $\psi$ particles with a decay rate $\Gamma_{\chi}$.
 Depending on the
couplings between particles, the decay process from $\phi$ to $\psi$ 
is so fast that all the energy stored in the oscillating field may turn
into $\psi$ particles. 
The $\psi$ particles then thermalize to complete the reheating
process. 
To determine the curvature perturbations produced during the
reheating process, it is important to consider the relation
$\rho_\chi\propto n_\chi$.\footnote{Here we followed the first paper in
Ref.\cite{IH-PR} and assumed instant decay (i.e., the preheat particles
decay at $m_\chi\simeq g\phi_c$ before $\phi$ turns around.
It is possible to construct inhomogeneous preheating models without the
instant decay. See for example Ref.\cite{IH-PR-withoutID}.}
Then, we obtain the
curvature perturbation $\zeta$ generated during 
preheating:
\begin{equation}
\zeta \equiv -\psi -H\frac{\delta \rho_\chi}{\dot{\rho_\chi}}
\simeq \beta \frac{\delta n_\chi}{n_\chi},
\end{equation}
where $\beta$ is a proportionality constant that depends on the
redshifting of the preheat particles, and in the last step we considered
a spatially flat gauge. 
To obtain an analytic estimate of the curvature
perturbation, from Eq.(\ref{preheat-n}) we find 
\begin{eqnarray}
\frac{\delta n_\chi}{n_\chi}&=&
\left[\frac{3}{2}+\frac{\pi g |\phi_{min}|^2}{|\dot{\phi}_{max}|}\right]
\frac{\delta |\dot{\phi}_{max}|}{|\dot{\phi}_{max}|}
-\frac{2\pi g |\phi_{min}|^2}{|\dot{\phi}_{max}|}
\frac{\delta |\phi_{min}|}{|\phi_{min}|}\nonumber\\
&\simeq& \frac{3}{2}\frac{\delta |\dot{\phi}_{max}|}{|\dot{\phi}_{max}|}.
\end{eqnarray}
Here, $\dot{\phi}_{max}$ is given by Eq.(\ref{dmax}), and the
long-wavelength inhomogeneities of $\sigma$ causes $\delta \omega$.
According to Ref.\cite{IH-PR}, multi-field inflation may lead to a
significant $\delta \phi_{min}$ and to $\phi_{min}\ne0$ at the minimum
of the oscillation, which causes
significant inhomogeneities of $n_\chi$.
However, considering single-field or symmetric multi-field potential,
the minimum of the oscillation trajectory is at $\phi=0$, 
and hence we may disregard terms proportional to $\phi_{min}=0$.
 The long-wavelength inhomogeneities of $\delta \omega$ then
lead to fluctuations of $\dot{\phi}_{max}$:
\begin{equation}
\delta \dot{\phi}_{max} \simeq -\frac{1}{2}\frac{\delta \omega}{\omega}
\dot{\phi}_{max},
\end{equation}
which eventually gives the curvature perturbation:
\begin{equation}
\zeta \simeq \beta \frac{\delta n_\chi}{n_\chi}\simeq 
\frac{3\beta}{4}\frac{\delta |\omega|}{|\omega|}.
\end{equation}
For the specific choice of $\omega$ given by
Eq.(\ref{choice-omega}), we find
\begin{equation}
\frac{\delta \omega}{\omega} \simeq \frac{\omega_\sigma}{\omega}
 \delta \sigma 
+\frac{1}{2}\frac{\omega_{\sigma\sigma}}{\omega}(\delta\sigma)^2
\simeq \alpha \frac{2\sigma}{M_*} \frac{\delta \sigma}{M_*}
+\left[
2\alpha^2 \frac{\sigma^2}{M_*^2} + \alpha 
\right]
\frac{(\delta\sigma)^2}{M_*^2}
\end{equation}
and the curvature perturbation is then
\begin{equation}
\zeta \simeq 
\frac{3\beta|\alpha|}{4}\left|
 \frac{2\sigma}{M_*} \frac{\delta \sigma}{M_*}
+\left\{
2\alpha \frac{\sigma^2}{M_*^2} + 1
\right\}
\frac{(\delta\sigma)^2}{M_*^2}\right|.
\end{equation}

The level of the non-Gaussianity is specified by the non-linear
parameter $f_{NL}$, which is defined by the Bardeen potential $\Phi$:
\begin{equation}
\Phi=\Phi_{Gaussian}+f_{NL}\Phi_{Gaussian}^2,
\end{equation}
which is connected to the curvature perturbation $\zeta$ through
\begin{equation}
\Phi=-\frac{3}{5}\zeta.
\end{equation}
If the first-order perturbation is dominantly generated by the usual
inflaton perturbation, the second-order perturbation that is generated
by the long-wavelength moduli inhomogeneities is not correlated to the
first-order perturbation. 
In this case, the estimate of the non-linear
parameter is given by \cite{unco-NG}: 
\begin{equation}
\frac{6}{5}f_{NL}\simeq \frac{1}{N_\phi^4}\left[
N_\sigma^2N_{\sigma\sigma}+N^3_{\sigma\sigma}{\cal P}_\sigma
\log(k_b L)
\right],
\end{equation}
where the curvature perturbation $\zeta$ is expanded by the 
$\delta N$ formalism as
\begin{equation}
\zeta \simeq N_\phi \delta \phi + N_\sigma\delta\sigma
+\frac{1}{2} N_{\phi\phi} \delta \phi^2 
+\frac{1}{2} N_{\sigma\sigma} \delta \sigma^2 + ...,
\end{equation}
and the perturbation can be separated into two parts:
\begin{equation}
\zeta = \zeta^{(\phi)}+\zeta^{(\sigma)}.
\end{equation}
Here $k_b\equiv$ min$\{k_i\}$ $(i=1,2,3)$ is the minimum wavevector of the
bispectrum and $L$ is the size of a box in which the perturbation is
defined.
The equation for the non-linear parameter can be simplified to 
obtain\cite{Lyth_and_Rod_NG} 
\begin{equation}
f_{NL}\simeq \left(\frac{1}{1300}
\frac{N_{\sigma\sigma}}{N_\phi^2}\right)^3.
\end{equation}
Assuming $\sigma \ll M_*$ at reheating, we obtain 
\begin{equation}
f_{NL} \simeq \left(10^{6} \times \alpha \beta 
\frac{H^2}{M_*^2}\right)^3,
\end{equation}
which becomes large as $H$ approaches the cut-off
scale $M_*$.
Although it depends on the parameters of the model, the result is very
interesting since the upper bound for $f_{NL}$ may apply a significant
upper bound on the inflation energy scale. 
The calculation suggests that the observation of the non-linear
parameter $f_{NL}$ may allow the inflation energy 
scale to be determined, dependent on the model parameters determined by
some other experiments.

\section{Conclusions and discussions}
In this paper, we considered reheating after inflation with a
non-standard kinetic term. 
We have shown that the difference in the
kinetic term in different Hubble patches inherited from the
long-wavelength moduli inhomogeneities $\delta \sigma$ causes generation
of a significant level of non-Gaussianity after inflation.

For completeness, we consider the tachyonic evolution of the light field
$\sigma$ during the oscillating phase. 
During the half-oscillation the oscillating field $\phi$ follows
$\dot{\phi}(t)= -\frac{m_\phi}{\sqrt{\omega}}\phi_0 
\sin \frac{m_\phi}{\sqrt{\omega}} t$, which
leads to the equation of motion for 
the light field $\sigma$: 
\begin{equation}
\ddot{\sigma}+3H \dot{\sigma}+W_\sigma-\frac{1}{2}\omega_\sigma 
\frac{m_\phi^2}{\omega}  \phi_0^2 \sin^2 m_\phi t =0.
\end{equation}
Omitting the term proportional to $\dot{\sigma}$ and disregarding the
variation of  $\omega$, 
the equation can be reduced to the well-known Mathieu equation for the
specific choice of $\omega$ given by Eq.(\ref{choice-omega}).
However, comparing the above equation with the standard preheating
equation for the original model\cite{instant-PR}, the coupling $g$ in
the standard interaction term $g^2 \phi^2\sigma^2/2$ is replaced by an
effective coupling  
\begin{equation}
g_{eff} \sim \alpha \frac{m_\phi^2}{M_*^2 \omega},
\end{equation}
which is very small if $m_\phi \ll M_*$.
We can also find a small slow-roll parameter $\epsilon^{eff}_{\sigma}\ll
1$ for $\sigma \ll M_*$.
Therefore, in this case the light field $\sigma$ is also 
slow-rolling in the oscillating phase.
Using a similar analysis for the initial condition $\sigma \ll M_*$, we
find that the fluctuation of the light 
field $\delta \sigma$ does not evolve exponentially during the oscillating
phase, even if $\eta_\sigma^{eff}$ is negative (i.e., tachyonic) during
this phase. 
Besides the possibility of exponential growth of the
long-wavelength fluctuation $\delta \sigma$, which may take place for
$\sigma \sim M_*$ and $\eta_\sigma^{eff}<0$, it is possible to consider
a more generic form of the effective action. 
In this paper, we considered a simple form of the potential assuming a
minimum at $\phi_{min}=0$.
However, in more generic cases we may consider a symmetry breaking
potential where the global minimum of the potential is not at the
origin.
Recently, preheating with a non-standard kinetic term and with a
symmetry breaking potential has been studied by Lachapelle et
al.\cite{nin-kin}, 
in which the oscillating field determines the coefficient $\omega$.
 It has been
shown that the non-standard kinetic term may give rise to a more
efficient preheating channel than the usual process caused by the
standard interaction term. 
It is possible to extend our arguments to
more generic actions based on some specific string models containing
non-standard kinetic terms that depend on several moduli fields. 
We can also extend our analysis to more generic initial conditions. 
In such cases, non-standard kinetic terms may give rise to more efficient
preheating than the standard interaction, as discussed by Lachapelle et
al., or they may cause tachyonic enhancement of $\delta \sigma$ during
oscillation. 
These effects may lead to a more significant level of
non-Gaussianity and to a more effective constraint for the effective
action. 
These complementary scenarios require numerical study, hence 
they will be considered in future works.

\section{Acknowledgment}
We wish to thank K.Shima for encouragement, and our colleagues at
Tokyo University for their kind hospitality.

\end{document}